\documentclass[conference]{IEEEtran}
\IEEEoverridecommandlockouts
\usepackage{cite}
\usepackage{amsmath,amssymb,amsfonts}
\usepackage{algorithmic}
\usepackage{graphicx}
\usepackage{multicol}
\usepackage{textcomp}
\usepackage{xcolor}
\usepackage[utf8]{inputenc} 
\usepackage[T1]{fontenc}    
\usepackage[hidelinks]{hyperref}       
\usepackage{url}            
\usepackage{booktabs}       
\usepackage{amsfonts}       
\usepackage{nicefrac}       
\usepackage{microtype}      
\usepackage{lipsum}
\usepackage{fancyhdr}       
\usepackage{graphicx,float}  
\def\BibTeX{{\rm B\kern-.05em{\sc i\kern-.025em b}\kern-.08em
    T\kern-.1667em\lower.7ex\hbox{E}\kern-.125emX}}
\begin{document}

\title{

ChatDiet: Empowering Personalized Nutrition-Oriented Food Recommender Chatbots through an LLM-Augmented Framework
}



\author{\IEEEauthorblockN{ Zhongqi Yang$^1$, Elahe Khatibi$^1$, Nitish Nagesh$^1$, Mahyar Abbasian$^1$, \\Iman Azimi$^1$, Ramesh Jain$^1$, and Amir M. Rahmani$^{1,2}$}

\textit{$^1$Department of Computer Science, University of California, Irvine}\\
\textit{$^2$School of Nursing, University of California, Irvine}\\

\IEEEauthorblockA{
\{zhongqy4, ekhatibi, nnagesh1, abbasiam, azimii, rcjain, a.rahmani\}@uci.edu}}



\maketitle

\begin{abstract}

The profound impact of food on health necessitates advanced nutrition-oriented food recommendation services. Conventional methods often lack the crucial elements of personalization, explainability, and interactivity. While Large Language Models (LLMs) bring interpretability and explainability, their standalone use falls short of achieving true personalization. 
In this paper, we introduce ChatDiet, a novel LLM-powered framework designed specifically for personalized nutrition-oriented food recommendation chatbots. ChatDiet integrates personal and population models, complemented by an orchestrator, to seamlessly retrieve and process pertinent information. 
The personal model leverages causal discovery and inference techniques to assess personalized nutritional effects for a specific user, whereas the population model provides generalized information on food nutritional content.
The orchestrator retrieves, synergizes and delivers the output of both models to the LLM, providing tailored food recommendations designed to support targeted health outcomes.
The result is a dynamic delivery of personalized and explainable food recommendations, tailored to individual user preferences. Our evaluation of ChatDiet includes a compelling case study, where we establish a causal personal model to estimate individual nutrition effects. Our assessments, including a food recommendation test showcasing a 92\% effectiveness rate, coupled with illustrative dialogue examples, underscore ChatDiet's strengths in explainability, personalization, and interactivity.

\end{abstract}

\begin{IEEEkeywords}
Large Language Model, Personalization, Explainability, Interactivity, Chatbots, Recommender Systems, Causal Reasoning, Nutrition, Food.
\end{IEEEkeywords}

\section{Introduction}

Food plays a pivotal role in our lives, exerting a profound impact on human health. Extensive studies affirm that nutrition not only contributes to overall well-being but also plays a crucial role in disease management, sleep regulation, mood modulation, and immune function enhancement~\cite{mathers2019paving,hu2002dietary,raut2018personalized}.
A compelling illustration of this connection lies in the influence of dietary choices on the gut microbiome. The Mediterranean diet serves as an exemplary model in this regard. Rich in fruits, vegetables, and cereals, and low in fat, this dietary approach emerges as instrumental in preserving and promoting gut health~\cite{Singh2017_gutMicrobiomeHumanHealth, Valdes2018_GutMicrobiotaNutritionHealth}.

This intricate relationship between diet and health has sparked a growing interest in leveraging technology for food recommendation services. These services provide guidance on diet plans, contributing to improved health outcomes.
Regular food recommendation services typically prioritize individual food preferences based on whether they like a particular food, with less emphasis on considering nutritional health aspects~\cite{min2019food}.
In contrast, nutrition-oriented food recommendations, as a form of food recommendation, recognize the significant impact of dietary patterns on individual health.

Numerous studies have contributed to the development of a variety of nutrition-oriented food recommendation methods~\cite{micha2017association,story2008creating,agapito2018dietos,iwendi2020realizing}. 
These approaches not only simplify the process of making dietary choices but also promote the adoption of sustainable and health-conscious habits among users.
The power of these recommendations lies in their ability to integrate the science of nutrition for population or sub-population level requirements, offering an intuitive and practical guide for individuals seeking to make informed dietary decisions.

Nonetheless, conventional nutrition-oriented food recommendation services have encountered limitations in comprehensively understanding the intricate interplay between individuals' health and well-being, encompassing physiological parameters, physical activity, and sleep quality, and aligning them with personalized nutritional needs.
More precisely, these systems frequently encounter challenges in tailoring their food suggestions to meet an individual's specific nutritional requirements. The inherent variability in how nutrition impacts individuals gives rise to concerns regarding the absence of genuine \textit{personalization}. Moreover, there is a prevalent reliance on population-based nutritional standards, potentially overlooking the nuanced individual differences in nutritional response.

Furthermore, current food recommendation services lack the essential element of \textit{Explainability}. These services often depend on machine learning models that function as "black boxes," rendering it challenging to elucidate the rationale behind their recommendations. Notably, nutrition-oriented food recommendation services grounded in population-based standards face difficulties in elucidating recommendations tailored to individualized nutrition requirements.
An additional substantial shortcoming lies in the absence of \textit{Interactivity}. These systems fall short in dynamically responding to user feedback, incorporating new preferences, and adjusting to changes in user status, such as temporary dietary choices or restrictions related to specific foods.

We posit that current advancements in Large Language Models (LLMs) present an opportunity to overcome these limitations.  LLMs have the capability to offer both interpretable and explainable recommendations, especially when harnessed in an interactive capacity such as acting as a chatbot~\cite{jeon2023large,birkun2023large}. However, it is essential to recognize that LLMs alone fall short in delivering the necessary degree of personalization required for nutrition-oriented food recommendations.  They lack the ability to integrate and leverage personal data into their analysis for recommendation and user interaction. Hence, a holistic framework is needed that utilizes LLMs for food recommendations while integrating diverse population-level and personal data sources.

In this paper, we present an LLM-powered framework for personalized nutrition-oriented food recommender chatbots named ChatDiet. 
We present an orchestrator, acting as a problem solver, to address food-related inquiries. This orchestrator analyzes input queries and extracts relevant information through engagement with personal and population models. 
Subsequently, responses are generated by feeding the information obtained to an LLM. 
 We utilize causal discovery and inference methods to extract a causal personal model from the user’s data, considering the impact of personal nutrition on health outcomes.
 We showcase the capabilities of the framework through a case study that includes health data from an individual collected over three years using wearable and mobile devices.  
We evaluate ChatDiet with a food recommendation effectiveness test to assess its ability to deliver personalized and explainable food recommendations. We also present examples of dialogues to indicate ChatDiet interactivity.

\section{Background and Related Work}

In this section, we first outline related work on nutrition-oriented food recommendations. We then provide an overview of LLMs implementations on recommendation tasks.

\subsection{Nutrition-oriented Food Recommendation}
Nutrition-oriented diet and food recommendation methods prioritize the nutritional content of foods when suggesting dietary choices. This shift from retail-focused models to nutrition-oriented diets in recommender systems aligns with a broader strategy toward preventive health~\cite{min2019food}. In the following section, we outline recent developments in diet and food recommendations, emphasizing the critical role of nutrition.

Recent studies have focused on customizing food recommendations to meet specific nutritional needs, particularly for the population with health conditions. For example, \cite{agapito2018dietos} leverages users' health status to suggest food beneficial for chronic kidney disease patients, while \cite{iwendi2020realizing}  integrates food features, biometric data, and disease-specific information to develop patient food recommendations. Additionally, \cite{raut2018personalized} uses a colony algorithm considering food information, population nutrition requirements, and daily activity to tailor recommendations for various groups of patients with diseases.
Another significant concern linked to diet, diabetes management, has been the subject of specialized research. Studies by \cite{sapri2019diet,zadeh2019personalized}  focus on food recommendations for diabetes patients. \cite{toledo2019food} explore a comprehensive approach that accounts for both preferences and nutritional needs, particularly in overweight and diabetic individuals. However, it only considers personal preferences but not personal nutrition requirements.

Extending beyond disease-specific recommendations, there are efforts to guide dietary choices in healthy populations. \cite{elsweiler2015towards}  propose meal plans balancing taste and nutrition, while \cite{ng2017personalized} introduce a toddler recipe system merging nutrition guidelines with user preferences. \cite{cioara2018expert,taweel2016service}  develop models for older adults, considering their unique nutritional needs.
A notable study by \cite{ge2015health} marks progress in calorie intake recommendations focusing on calorie count. The need for nutrition-oriented food recommendations is further demonstrated by advancements in food knowledge graphs and machine learning~\cite{Min2021_foodKGReview, Gharibi2020_foodKG_ML}. Incorporating factors like physiology, genetics \cite{Sikka2019LearningIFood2019Challenge}, and microbiomes makes dietary advice more relevant to a certain population.

These studies highlight the importance of integrating contexts like disease status, daily activity, and biometric data. 
Nevertheless, though users may share similar health objectives, achieving those objectives may require different dietary choices based on individual variations. This stems from the fact that factors such as genetics, behaviors, lifestyles, and personal preferences contribute to variations in how people respond to specific foods. Thereby a personalized approach to nutritional recommendation acknowledges the need to tailor dietary plans to the unique requirements of each individual. 
 
Furthermore, the explainability in these models is limited, with a notable exception being \cite{iwendi2020realizing}, which provides insights into feature importance. However, the machine learning models operate as "black boxes," which means it is often challenging to understand the reasons behind their recommendations. The current nutrition-oriented food recommendation services that rely on population-based standards struggle to provide explanations tailored to an individual's specific nutritional needs.
 
Interactive features, including adaptively responding to user feedback, are also generally lacking. Food recommendation systems should be able to adjust based on user feedback, accommodate new preferences, and adapt to changes in user status, including temporary dietary choices or restrictions related to specific foods.

\subsection{Exploiting LLMs for Recommendation Tasks}

Pre-trained LLMs exhibit remarkable adaptability across diverse recommendation tasks~\cite{cui2022m6rec}, particularly when personalized historical data becomes a key component. The landscape of LLM research in recommendation scenarios can be categorized based on the LLM's role as either the central recommendation model or a facilitator of recommendation tasks.

Numerous studies incorporate LLMs as central recommendation models or integral components. To harness LLMs' natural language prowess, data transformation into textual formats is pivotal without training~\cite{huang2023chatgpt}. Prompt engineering, a prevalent technique, converts structured personal data into text-based inputs. This process entails designing informative prompts, converting interactions and attributes into text, and forming coherent inputs for LLM processing~\cite{liu2023pre}.

In a pilot study presented in \cite{zhang2021language}, recommendation tasks are transformed into multi-token close tasks through prompts, aimed at mitigating zero-shot learning challenges. Other works, such as~\cite{liu2023chatgpt} and~\cite{wang2023zero}, employ diverse prompts for recommendation scenarios. \cite{geng2022recommendation} introduces task-specific prompt templates with inserted personal data, while~\cite{dai2023uncovering} reformulates ranking tasks into prompts. LLM-rec~\cite{lyu2023llm} combines strategies for personalized content recommendation, and Tallrec~\cite{bao2023tallrec} transcribes user history to instruct LLMs. \cite{hou2023large} employs LLMs as item rankers based on interaction histories. In contrast, \cite{zhang2023recommendation} adopts instruction tuning, aligning LLMs with recommendations through templates that incorporate user interactions. A different approach emerges in \cite{li2023prompt}, using soft prompts generated from user and item embeddings. These generated aspects act as intermediaries within LLM recommendation systems, offering an avenue to integrate abstract personal data representations into LLM-based recommendations.

Existing LLM-based methodologies face limitations in directly using personal data, including historical user interactions and preferences, as textual input.
They often miss out on employing models or encoders that can extract meaningful representations from this data.
In an attempt to tackle this challenge, a recent study \cite{li2023prompt} proposes an approach that utilizes abstract embeddings as intermediaries to bridge the gap. However, it is important to note that this approach compromises explainability because abstract embeddings can be opaque in nature.

In the domain of nutrition-oriented food recommendations, individual nutritional impacts are not directly observable and require inference from the existing personal data. This situation highlights a methodological shortcoming of sole LLM: the inability to adapt effectively when personal data, unlike straightforward historical interactions, is implicitly relevant for recommendation purposes.
While retrieval-augmented generation~\cite{lewis2021retrievalaugmented} offers a method for enabling LLMs to incorporate external information by retrieving relevant data, in the context of this study, there are no pre-existing resources available for retrieval. 
This raises a need for a more comprehensive exploration of the Personalization and Explainability aspects within the food recommendation systems.
Incorporating abstract data on physiological parameters and health information, alongside nutritional needs, into LLMs while maintaining explainability is a critical aspect of delivering personalized dietary guidance.

\section{ChatDiet: An LLM-Based Framework for Nutrition-oriented Food Recommendation}

\begin{figure*}[t!]
    \centering
    \includegraphics[width=0.8\textwidth]{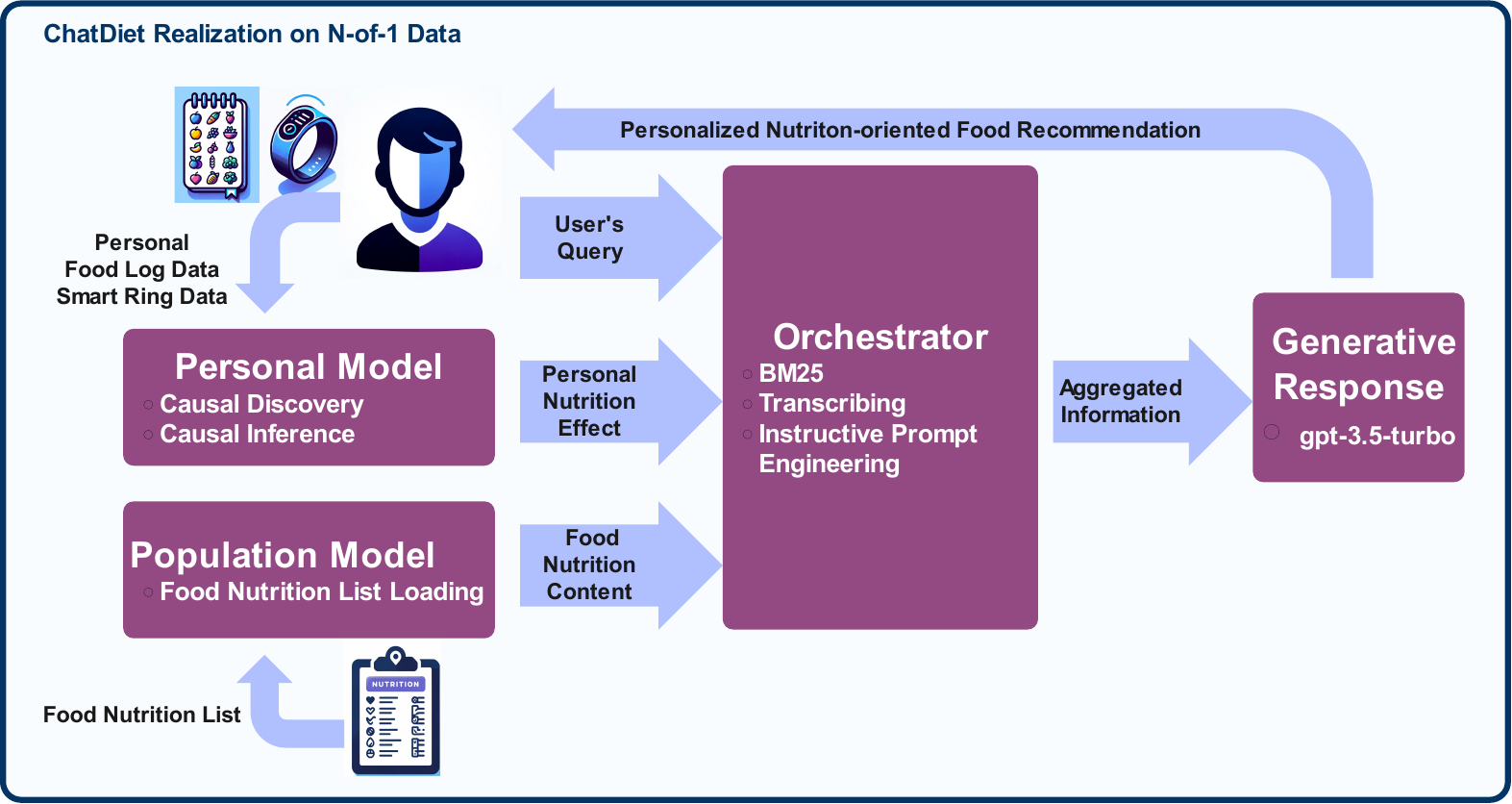}
    \caption{Overview of ChatDiet Realization on N-of-1 Data.}
    \label{fig:chatbot}
\end{figure*}

In this section, we introduce ChatDiet, a framework for personalized nutrition-oriented food recommendation chatbots.
This framework utilizes LLMs to effectively incorporate not only population knowledge but also individual-specific data through augmented models. It includes an Orchestrator that interacts with personal and population models to extract relevant information based on users' inquiries. It then sends the aggregated information to an LLM to be integrated with the LLM's internal knowledge and offer interactions with users. Figure~\ref{fig:chatbot} indicates an overview of the framework.

To clarify the functionality and definitions of the framework, we present and exemplify different components of ChatDiet via a case study. In the following, we initially delve into our case study, focusing on mHealth system leveraging wearable and mobile-based data logging.

\subsection{Case Study} \label{sec:casestudy}

Our case study includes data collected from an individual's dietary habits, health metrics, and other relevant information during three years. We focus on the intricate connection between dietary habits, physical activity, sleep, and health outcomes, aiming to leverage it to effectively address the task of providing personalized conversational food recommendations.

\subsubsection{Dataset}

\textbf{N-of-1 Dataset:}
The N-of-1 dataset was collected over a period of more than three years from a participant using multiple devices. The user was recruited to wear an Oura ring~\cite{oura} between mid-April 2020 to mid-June 2022 to gather his/her sleep quality~\cite{mehrabadi2020sleep}, physical activity~\cite{niela2022comparison}, skin temperature, heart rate, and heart rate variability (HRV) \cite{cao2022accuracy}. Additionally, a non-invasive smart weighing scale, Arboleaf~\cite{arboMadhu2018Personalized}, was used from the first week of January 2020 to the first week of June 2022 to measure body fat and body composition parameters, including body weight, body mass index (BMI), visceral fat, bone weight, among others.
 
To track meals consumed during breakfast, lunch, dinner, and snack times, the user used the Cronometer food-logging mobile application~\cite{cronom} from July 1, 2019, to mid-June 2022. We use the calorie consumption (in kCal) along with the macronutrients and macronutrients quantities parsed from the app as a result of the user's meal input to obtain the calorie and nutrition intake. We also integrate the individual's health metrics related to height, weight, and blood pressure stored in the Apple Health Kit~\cite{applhealth} from mid-October 2018 to mid-June 2022. This allows us to incorporate lifestyle metrics, thereby creating a data-driven persona of a user's overall health and well-being.

\textbf{Synthetic Dataset:} 
As it is challenging to evaluate ChatDiet on N-of-1 data, to quantitatively assess ChatDiet with a larger sample size, we augment the dataset by generating synthetic samples. 
This expansion follows a structure similar to that of the collected dataset. Specifically, we have synthesized 365 days of data for a simulated cohort comprising one hundred augmented participants. 
Each augmented participant is associated with a personalized causal graph that links nutritional factors to health outcomes that are identical to the graph of N-of-1 data. These graphs are generated with random individual treatment effects.

To simulate daily nutrition intake patterns for each participant, we have assigned random values representing their nutritional consumption and health outcomes. 
These values are drawn from a Gaussian distribution that has been fitted using the N-of-1 data as a reference to minimize deviations from factual observations.
Subsequently, for each synthetic participant, we estimate the individual treatment effects of nutrition on each health outcome.
These personal nutrition effects are determined based on the participant's nutrition intake, the causal graph specific to that individual, and the corresponding health outcome values. This process enables us to evaluate the performance of our framework within a satisfactory sample size.

\subsection{Architecture}
The ChatDiet framework includes 4 major components: Orchestrator, Personal Model, Population Model, and LLM (see Figure \ref{fig:chatbot}). In the following, we briefly outline the components, followed by their implementation using the case study.

\subsubsection{Orchestrator}
We introduce the Orchestrator within our ChatDiet framework inspired by the Conversational Health Agents framework (openCHA) proposed in \cite{abbasian2023conversational}. The Orchestrator plays a central role in processing and enhancing the effectiveness of ChatDiet's food recommendations as it is responsible for filtering, fusing, and transcribing the outputs generated by personal and population Models and user queries. Within the ChatDiet framework, the Orchestrator performs three main processing tasks:

\textit{Retrieving}: It retrieves the most relevant information from the Personal Model and Population Models based on the user's specific query. This retrieval process ensures that only pertinent data is considered for generating responses. For instance, if a user's health data contains extensive information, but only details related to specific health outcomes are required, the Orchestrator is responsible for filtering and retrieving only that subset of data.

\textit{Transcribing}: It transforms non-textual information into a textual format, enabling the optimal utilization of input data across diverse modalities (e.g., time series). These converted inputs are then fed into LLMs, designed to primarily handle textual inputs. This capability proves crucial, especially when confronted with data formats such as spreadsheets or other non-textual structures that the LLM cannot directly interpret. 

\textit{Prompt Engineering}: It performs a prompt engineering function to instruct the LLM effectively. It ensures that the LLM understands its role and context in the conversation. Additionally, when personal food preferences and nutrition effects on health are provided, the Orchestrator supplies prompts that instruct the language model to adhere to the provided information rather than generating unrelated or hallucinatory responses. This step enhances the quality and relevance of the chatbot's interactions.

\textbf{The case study:} For our case study, we build the Orchestrator, equipped with the Best Match 25 (BM25) retrieval algorithm~\cite{robertson2009probabilistic}), instructive prompts, Chain-of-Thought prompt engineering methods~\cite{wei2022chain}, and a transcribing mechanism. These components collectively facilitate retrieving and transforming personal nutrition effects and food nutrition data into suitable text formats.

We implement a two-stage retrieval process, aimed at only selecting relevant nutrition effects and foods that align with the user's query. The first stage is to retrieve relevant nutrition effects and food pertinent to the user's query. For instance, if the user's query involves enhancing deep sleep, we leverage the BM25~\cite{robertson2009probabilistic}) algorithm to retrieve nutrients that significantly impact deep sleep. Subsequently, the second stage focuses on retrieving food items that contain the most relevant nutrients. This is executed by ranking all foods based on their nutrition content per calorie, subsequently selecting the top 10 entries. 

Consequently, the causal impact of each nutrient on distinct health outcomes was encapsulated within this structure. 
For instance, utilizing the estimated individual treatment effect (ITE), the causal effect of vitamin B1 on the duration of deep sleep was captured as \textit{"The effect of B1 (Thiamine) (mg) on Deep Sleep Duration: 14.3 per unit."} The food ingredients are encoded as \textit{"Beet Greens, Cooked per cup contains 38.8kcal of Energy, 128.3g of Water, 0.16mg of B1, 0.41mg of B2."} This approach ensures explainability, accessibility, and consistency in presenting the complex causal relationships to the LLM.

We conduct prompt engineering in Orchestor based on instruction template~\cite{hou2023large} and the concept of engagement-guided instructions~\cite{lyu2023llm}. Leveraging the personal nutrition effects and food ingredients dictionary, we construct the instruction prompt as follows: \textit{"Please provide a food recommendation based exclusively on the nutrition effects and the provided list of food ingredients. Your recommendations must strictly adhere to the listed foods. You cannot generate foods that are not in the given food list. The relevant nutrition effects are as follows: [nutrition effects]. The foods' ingredients are as follows: [food ingredients dictionary]."}.

Moreover, we utilize the Zero-Shot Chain-of-Thought method~\cite{wei2022chain} to enhance user understanding and confidence in the recommendation. The Zero-Shot Chain-of-Thought method elevates accuracy and clarity by constructing coherent response sequences that logically build upon one another. To this end, the orchestrator appends the explicit instruction \textit{"Explain your recommendation step by step"} to the instruction prompt.
This not only ensures alignment with personal nutrition effects but also facilitates detailed explanations for food recommendations.

\subsubsection{Personal Model} The Personal Model incorporates a range of personal data into the recommendation process. In the context of ChatDiet, personal data encompasses unique individual-specific information, including personal food preference ratings, dietary history, electronic health records, and physiological signals gathered from wearable devices. It might include various data formats, extracting representations, embeddings, or textual descriptions.

An illustrative example of a Personal Model is the Biological Personal Food Model (B-PFM) and Preferential Personal Food Model (P-PFM) as introduced in \cite{rostami2020personal}. These models are designed to address how food items can effectively meet nutritional needs aligned with specific goals for an individual. They prioritize the individual's health state, considering their nutritional requirements, taste preferences, and biological responses to various foods.
 
\textbf{The case study:} We implement ChatDiet's Personal Model using causal discovery and causal inference methods to determine the causal effects of nutrition on health outcomes.
Causal discovery and causal inference are two main fundamental techniques in understanding the cause-and-effect relationships that govern various phenomena~\cite{feng2023causal}. 
These methods go beyond mere correlation by providing insights into how changes in one variable can lead to changes in another, allowing us to uncover the underlying mechanisms that drive observed outcomes. 

In our study, we leverage a novel NN-based causal discovery algorithm called Structural Agnostic Modeling (SAM) \cite{kalainathan2022structural}. SAM learns causal generative models in an adversarial manner. It aims to combine the advantages of exploiting conditional independence relations and leveraging distributional data asymmetries, achieving a trade-off between model complexity and data fitting.

After obtaining the output from the causal discovery algorithms, our focus shifts to estimating the causal impacts based on the causal graph, achieved through the application of inference methods. These methods leverage statistical and computational techniques, allowing researchers to infer the impact of one variable on another. The ATE~\cite{shalit2017estimating, yao2021survey}, a key concept in causal inference, quantifies the average impact of a treatment or intervention on an outcome. 

To enhance the personal model's ability to understand the complex connections between diet and well-being, we leverage insights from ~\cite{nagesh2023towards} and apply causal discovery and causal inference techniques. These methods help us to assess the causal effect of different nutrients on health outcomes.

We find that our final causal graph contains a mediator path, and therefore, we use the mediator analysis to compute the true causal estimation. Briefly, mediator analysis in causality involves investigating the intermediate processes through which an independent variable influences a dependent variable \cite{pearl2014interpretation}. Statistical methods, such as regression or structural equation modeling, are employed to assess the direct effect of the independent variable on the dependent variable and the indirect effect mediated through the intermediate variable. Mediator analysis is crucial for illustrating the underlying processes that contribute to observed causal relationships, providing insights into the pathways through which variables exert their influence in a causal chain \cite{pearl2012mediation}.

For the causal discovery, we employed the Causal Discovery Tool (CDT) library~\cite{kalainathan2019causal} on personal data to generate a causal graph depicting the relationships between daily nutritional intake and subsequent day's health outcome variables.
We then estimate the causal effects using the DoWhy library~\cite{sharma2020dowhy} by leveraging both personal data and the previously constructed causal graph. 

\subsection{Population Model}

A Population Model encompasses information that is not specific to a person but rather pertains to a larger group or population (in contrast to a Personal Model). 
Such population information is necessary to ensure that personal food recommendations are grounded in broader dietary trends and health norms. 
For food recommendation, population knowledge consists of a wide array of data, including but not limited to food knowledge graphs~\cite{haussmann2019foodkg}, general nutritional standards, public dietary guidelines~\cite{zeraatkar2019evidence}, and nutritional and supplement facts \cite{rossi2023alignment, cowan2023narrative}. The extracted information from the Population Model can take various forms, including text or non-text data, and is employed in downstream components within the system. This allows the ChatDiet framework to leverage more generalized population knowledge to enhance its recommendations.

\textbf{The case study:} We introduce the Population Model as a food knowledge loader function.
Its primary role is to select and load food’s nutrition content from the comprehensive food database created by the Cronometer Food Logger~\cite{cronom}, which details each food item's ingredients, nutritional content, and sensory properties like texture, smell, and taste. There are also other databases with APIs, such as Nutritionix~\cite{nutritionx}, which can be used for this purpose. The Population Model streamlines the process by directly selecting and loading this nutrition information into the Orchestrator, eliminating the need for complex data processing steps such as embedding extraction.

One notable advantage of this approach is the seamless accessibility of data, as the population model can effortlessly load this information without the need for additional embedding extraction processes. This efficient data retrieval ensures that ChatDiet can tap into a rich source of general food nutrition knowledge, enhancing the chatbot's ability to provide informative and contextually relevant recommendations.

\subsection{Generative Response}
The Generative Response is a key component in ChatDiet that leverages the language model to process the curated text from the Orchestrator. 
The objective of the Generative Response is to generate personalized, nutrition-oriented food recommendations by leveraging the personal context and population knowledge, along with the user’s query processed by the Orchestrator.

\textbf{The case study:} We employ the gpt-3.5-turbo \cite{brown2020language} Language Model. The model is responsible for processing the data received from the Orchestrator and generating responses.

\section{ChatDiet Evaluation}

In the following, we outline the results of a food recommendation effectiveness test and examples showcasing ChatDiet's Explanability, Personalization, and Interactivity. While not exhaustive, this evaluation demonstrates the platform's promising potential for personalized recommendations. However, a comprehensive personalization assessment in the field of LLM-based food recommendation depends on the development of standardized benchmarks and evaluation metrics, which are currently unavailable.
A concise demonstration video showcasing the ChatDiet framework implemented on openCHA is available at~\cite{ChatDietDemoVideo}. 
\subsection{Personal Model Results}
The primary objective of ChatDiet is to provide food recommendations based on the personal nutrition effect guided by the Personal Model.
We begin by presenting the outcomes derived from the Personal Model, as utilized by the subsequent Orchestrator and Response Generator. The key findings from this Personal Model, centered on causality, highlight the complex interplay between dietary consumption and personalized health responses. To accomplish this, we initially construct a causal graph using causal discovery techniques. Following that, we determine the individual treatment effects for each individual through causal inference methodologies.

The causal graph is shown in Figure~\ref{fig:causalgraph} in the Appendix section.
We observe significant correlations between nutritional intake and various physical health indicators, including activity levels, sleep patterns, and heart health. 
For example, certain nutrients have demonstrated a positive association with enhanced sleep efficiency, while others exhibit links to physical activity levels and heart rate measurements.
Notably, the graph reveals the complexity of interactions between macronutrients such as proteins, fats, and carbohydrates, and a wide range of health metrics. 
For example, the intake of specific types of fats, such as omega-3 and omega-6 fatty acids, intricately influences factors like heart rate variability and energy levels. 

In the causal inference results, we observe notable trends in the effects of micronutrients on various health outcomes. Specifically, micronutrients, including Vitamin D, Iron, and Magnesium, demonstrate significant effects on sleep quality, resting heart rate, etc. We can also observe the influence of certain macronutrients, including omega-3 fatty acids, and carbohydrate show a significant effect on health outcomes, including HRV, sleep efficiency, activity burn, etc. This underlines the importance of not only the quantity but also the quality of the nutrients in an individual’s diet.
These results are instrumental for the following Orchestrator and Response Generator. They enable the Orchestrator to access and process personal nutrition effects and empower the Response Generator LLM to craft personalized, nutrition-oriented recommendations.

\subsection{Quantitative Validation of Effectiveness}

We first show that the ChatDiet can efficiently provide valid food suggestions with explanations. To this end, we conduct a quantitative evaluation to assess the efficiency of recommendations provided by ChatDiet, while considering explanations and personal data. This evaluation is based on whether the recommendations were aligned with the user's query and if the accompanying explanations are consistent with the extracted personal effects. 

We selected HRV, overall sleep quality, REM sleep duration, and deep sleep duration as the target health outcomes. Then, the user queries were constructed in the format of 'Please suggest me a food to increase [target health outcome].' With four target health outcomes and 100 synthetic participants, we generated a total of 400 queries and responses from ChatDiet. We manually assessed the explanation provided in the response to ensure it included information on the food recommendation's nutrition content and its associated nutrition effects. 
If the presented nutrition effects align with the individual's personal nutrition requirements, we classify the recommendation as 'correct.' 
For instance, if Chatdiet outlines the rationale behind recommending Acai berries by stating that this food is rich in Vitamin E, and that Vitamin E has an effect on extending deep sleep duration by 3.3408 per unit, we then manually verify the estimated nutritional impact of Vitamin E. If both assessments align, we classify this recommendation as 'correct'.
The recommendation effectiveness ratio (RER) of ChatDiet is then calculated as the proportion of 'correct' recommendations. 
The results of this evaluation are presented in Table~\ref{tab:explanationacc}.

\begin{table}[htbp]
    \caption{Accuracy of Personalized Food Recommendations and Explanations}

\begin{center}
    \begin{tabular}{|c|c|}
    \hline
         & RER
         \\
          \hline
          Recommendations regarding  \textbf{HRV} &0.95  \\
          \hline
 Recommendations regarding  \textbf{Overall Sleep Quality}& 0.93\\\hline
 Recommendations regarding  \textbf{REM Sleep Duration}& 0.85\\\hline
 Recommendations regarding \textbf{Deep Sleep Duration}& 0.95\\\hline
    \end{tabular}
    \label{tab:explanationacc}
    \end{center}
\end{table}

\subsection{ChatDiet's Explainability Demonstration}

A pivotal characteristic of ChatDiet lies in its emphasis on explainability, which revolves around its ability to show the underlying logic and decision-making process behind recommendations. 
This attribute holds true for ChatDiet, as it employs personalized nutrition effects to expound upon its dietary recommendations in relation to the health outcomes of interest.

\begin{figure}[h]
    \centering
    \includegraphics[width=0.5\textwidth]{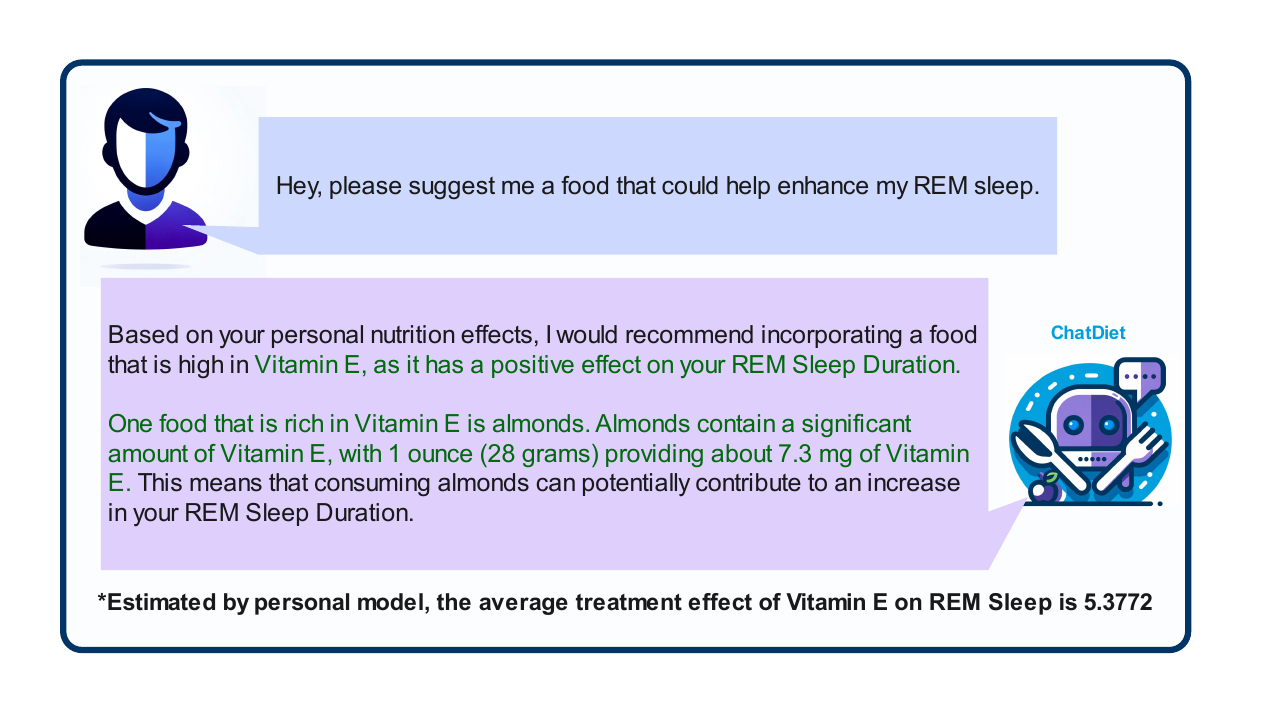}
    \caption{Food recommendation and explanation based on nutrition effect from ChatDiet.}

    \label{fig:explainability}
\end{figure}

In Figure~\ref{fig:explainability}, we present an example of ChatDiet's interaction dynamics. These showcases underline how the system's recommendations amalgamate personalized nutrition effects gleaned from the personal model, alongside inherent food-related expertise embedded in the pre-trained LLM's weights. In this example, if a user aims to improve their REM sleep duration, ChatDiet suggests foods abundant in Vitamin E, Valine, and Protein based on the user's personal model. It therefore recommends foods such as Almonds, which are rich in Vitamin E. 
Notably, ChatDiet substantiates this recommendation using personal nutrition effects, as all three of these nutrients exhibit positive impacts. 
This approach effectively connects individual dietary requirements with the broader nutritional expertise embedded within the LLM. 

\subsection{ChatDiet's Personalization Demonstration}
Personalization denotes the system's prowess in customizing food recommendations to align precisely with an individual's distinct nutrition effects. This personalized strategy aims to furnish dietary suggestions that resonate with the user's physiological profile, moving beyond population nutrition knowledge.

\begin{figure}[h]
    \centering
    \includegraphics[width=0.5\textwidth]{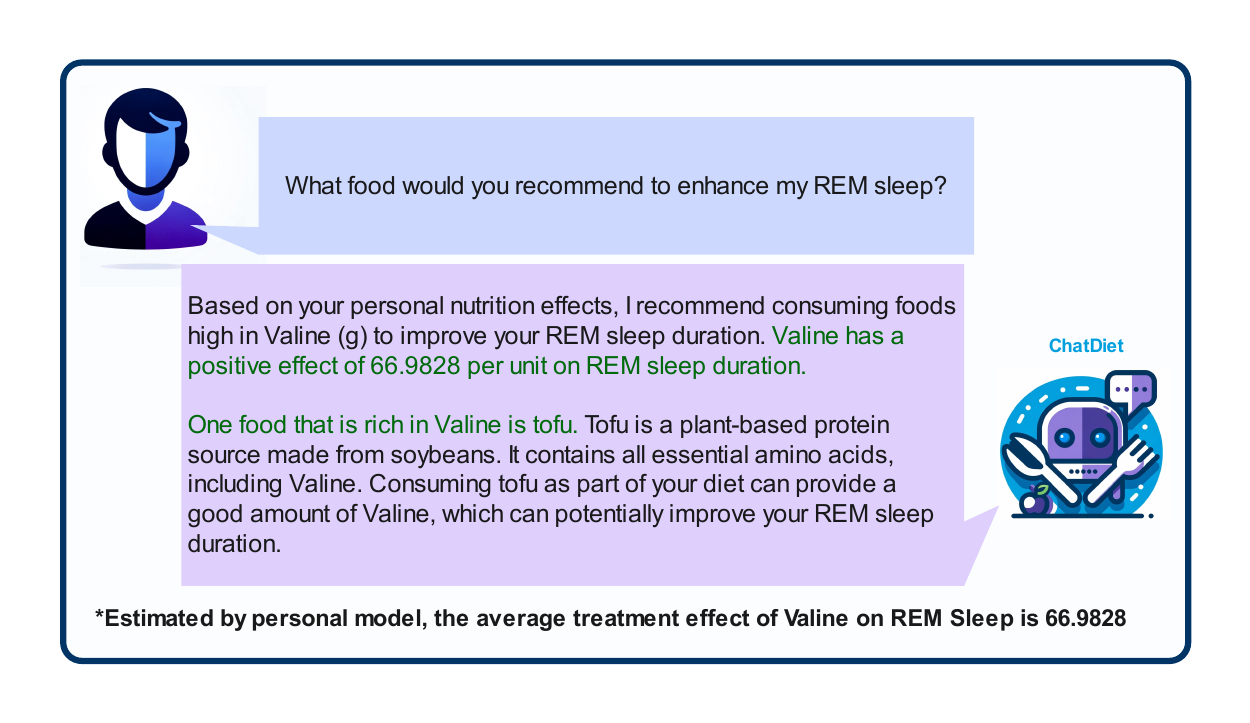}
    \vspace{-5mm}
    \caption{Personalized Feedback from ChatDiet.}
    \label{fig:personalization1}
\end{figure}

\begin{figure}[h]
    \centering
    \includegraphics[width=0.5\textwidth]{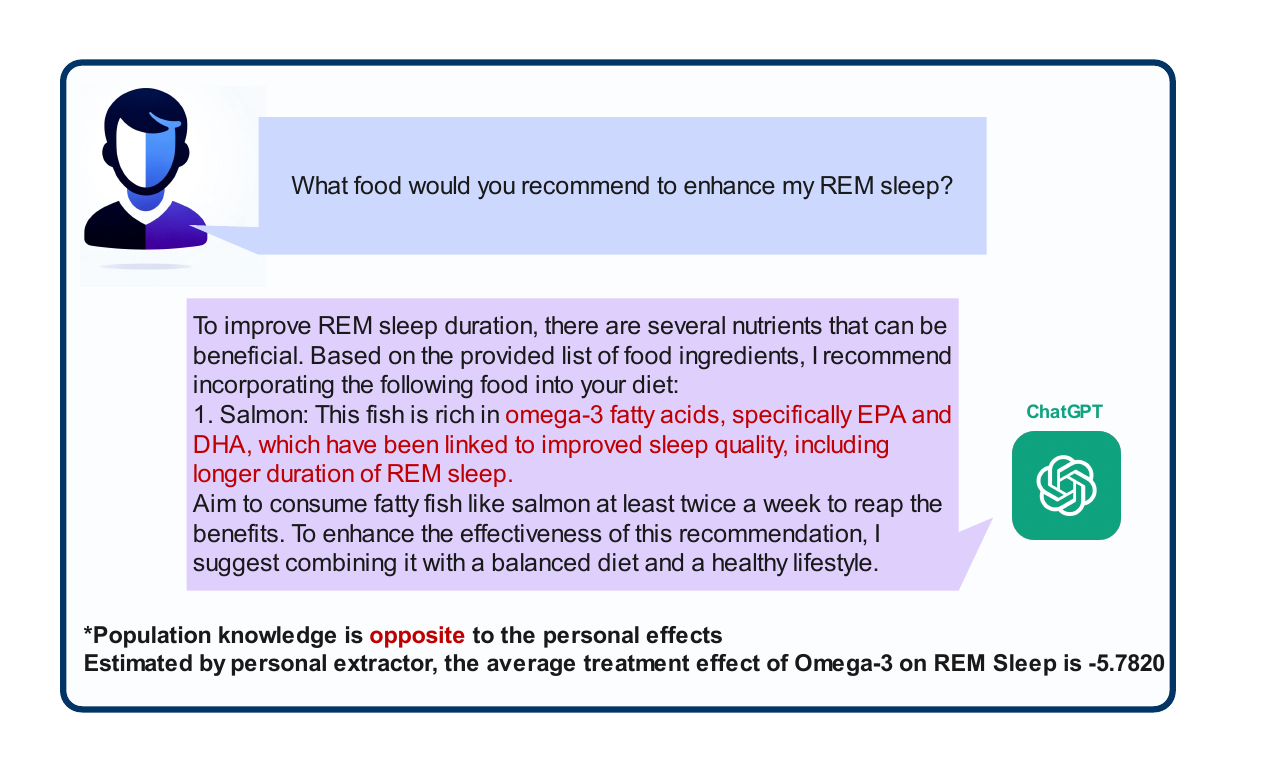}
    \caption{Non-personalized Feedback from ChatGPT.}
    \label{fig:personalization2}
\end{figure}

Personalization is showcased through the comparisons depicted in Figures~\ref{fig:personalization1}~and~\ref{fig:personalization2}. This graphical representation exhibits how ChatDiet transcends dependence solely on population-based knowledge and effectively integrates personal nutrition effects into its recommendations.
A compelling case study arises when we compare ChatDiet's recommendations both with and without the incorporation of personal data.

The system's competence in suggesting Salmon, typically linked to the favorable effects of Omega-3 on REM sleep, acquires a new dimension with the integration of an individual's personal nutrition effect.  ChatDiet demonstrates its advanced rationale by not only acknowledging the adverse impact of Omega-3 but also assessing the constructive influence of protein content. This multi-model methodology underscores ChatDiet's proficiency in crafting suggestions that meticulously consider an individual's unique nutritional nuances. It is noteworthy that this personalized approach stands in contrast to the broader population knowledge, suggesting the positive impact of Omega-3, further emphasizing ChatDiet's capability to adapt to personal needs.

By personalizing recommendations based on personal nutrition effects, ChatDiet not only enhances the relevance of its suggestions but also amplifies its potential to impact the user's health and well-being positively. This tailored approach aligns dietary choices more closely with the user's physiological requirements, manifesting a more meaningful and effective recommendation experience.


\subsection{ChatDiet's Interactivity Demonstration}

ChatDiet's strength in interactivity stands out due to LLM's capabilities. Beyond the initial responses, ChatDiet engages in dynamic conversations effectively, thereby addressing the challenge of accommodating evolving user preferences in real time. This dynamic feature extends beyond static suggestions, as ChatDiet shows proficiency in responding to follow-up questions, allowing users to seek insights, make adjustments, or express their preferences interactively.

\begin{figure}[h]
    \centering
    \includegraphics[width=0.5\textwidth]{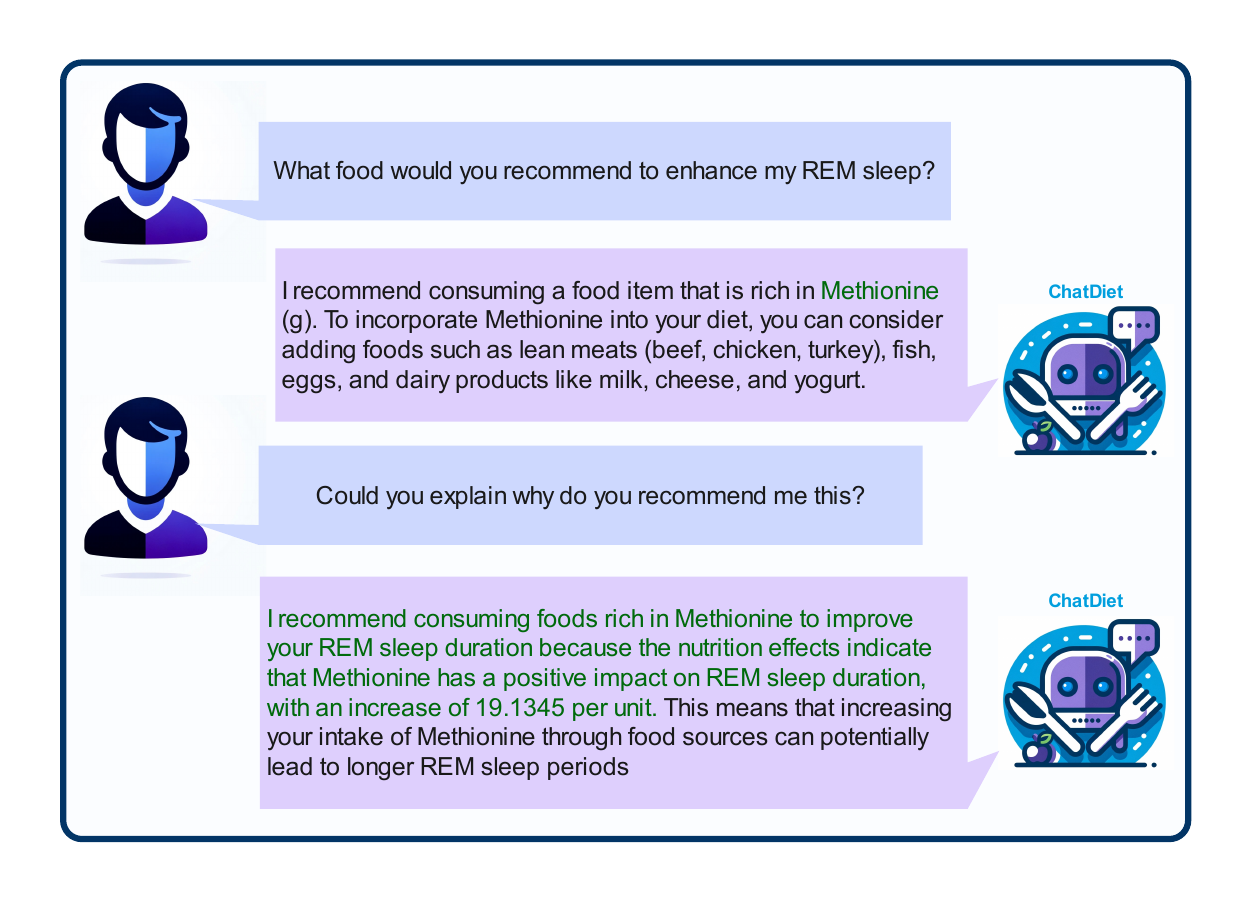}
    \caption{ChatDiet responds the follow-up question asking for further explanation.}
    \label{fig:interact1}
\end{figure}

\begin{figure}[h]
    \centering
    \includegraphics[width=0.5\textwidth]{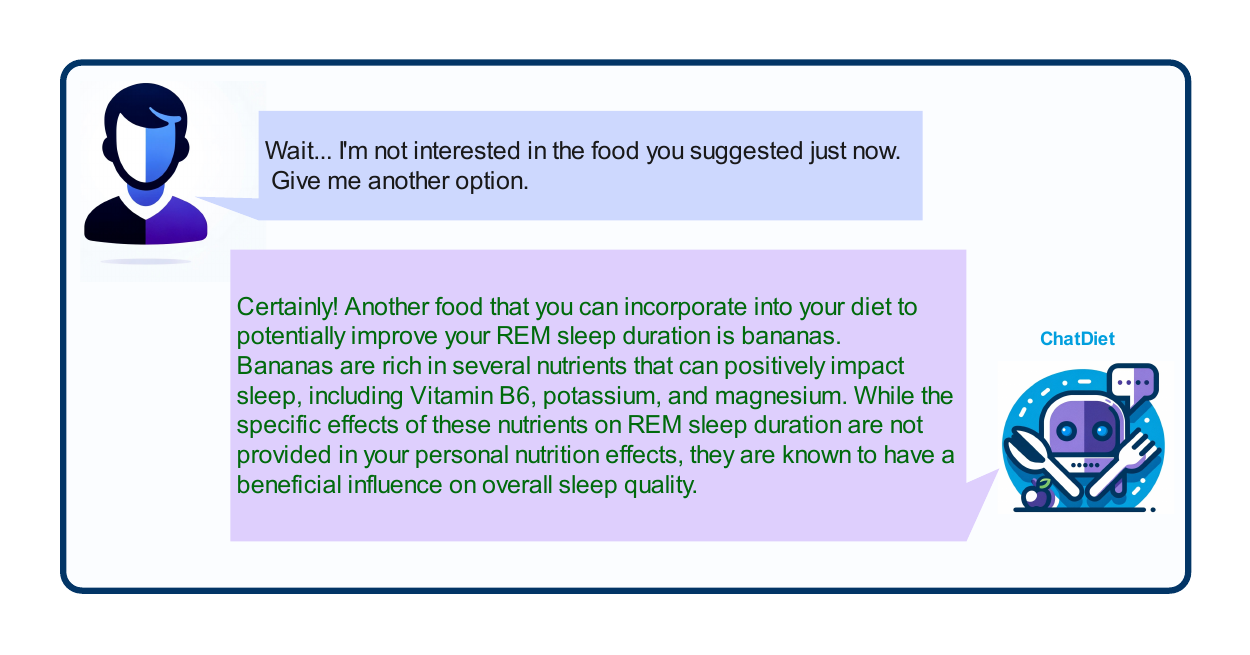}
    \caption{ChatDiet responds the follow-up question regarding altered preferences.}
    \label{fig:interact2}
\end{figure}

Figures~\ref{fig:interact1} and \ref{fig:interact2} illustrate ChatDiet's pronounced interactivity, particularly evident in its capacity to engage in follow-up questions. In instances where the initial response from ChatDiet lacks a detailed explanation, users have the option to pose further queries for additional clarification, thereby enabling a dynamic and interactive dialogue. The second exemplar showcases ChatDiet's ability to adapt; when a user expresses a lack of interest in the suggested food item, ChatDiet promptly offers alternative options. 
These demonstrations effectively underscore the extent to which ChatDiet successfully achieves and showcases interactivity within its recommendation process.


\section{Discussion}

The implementation of ChatDiet has indicated successes in explainability within the nutrition-oriented food recommendation. 
Several example dialogues have shown that ChatDiet is distinguished through its strong emphasis on explainability. This key feature enables users to gain clear insights into the rationale behind its recommendations. 
We show that by utilizing personalized nutrition effects, ChatDiet demonstrates dietary suggestions in the context of health outcomes, ensuring a transparent decision-making process.

Personalization in ChatDiet is shown by its ability to customize food recommendations to match an individual's unique nutrition effects. 
This personalized approach ensures dietary suggestions align with the user's specific physiological profile, going beyond population nutrition knowledge. 
By utilizing this, ChatDiet enhances relevance, promoting better health outcomes through tailored dietary choices.

ChatDiet's interactivity is evident through its ability to respond to users' follow-up questions effectively. The adaptability ensures recommendations remain pertinent as user preferences evolve, contributing to user control. Its dynamic interaction capability extends the initial responses, which is critical for engaging users in dynamic conversations. This enables users to seek insights, make real-time adjustments, or express their evolving preferences, ultimately fostering a highly interactive and user-centric experience.

However, ChatDiet faces certain limitations. Notably, ChatDiet's personalized food suggestions are confined to queries related to the factors available in the dataset, as the data is derived from a single subject and specific health outcomes from a limited set of smart devices. In cases where a user's query lacks explicit factor names, while the LLM can comprehend the user's intent and perform the task, occasional inaccuracies may arise, potentially leading to nonsensical recommendations. Our future work will focus on incorporating counterfactual analysis of user preferences under hypothetical scenarios to predict their future engagement and generate personalized recommendations based on these simulations. This will enable us to move beyond analyzing past behavior and provide more accurate and effective services for our users 

\begin{figure}[h]
    \centering
    \includegraphics[width=0.5\textwidth]{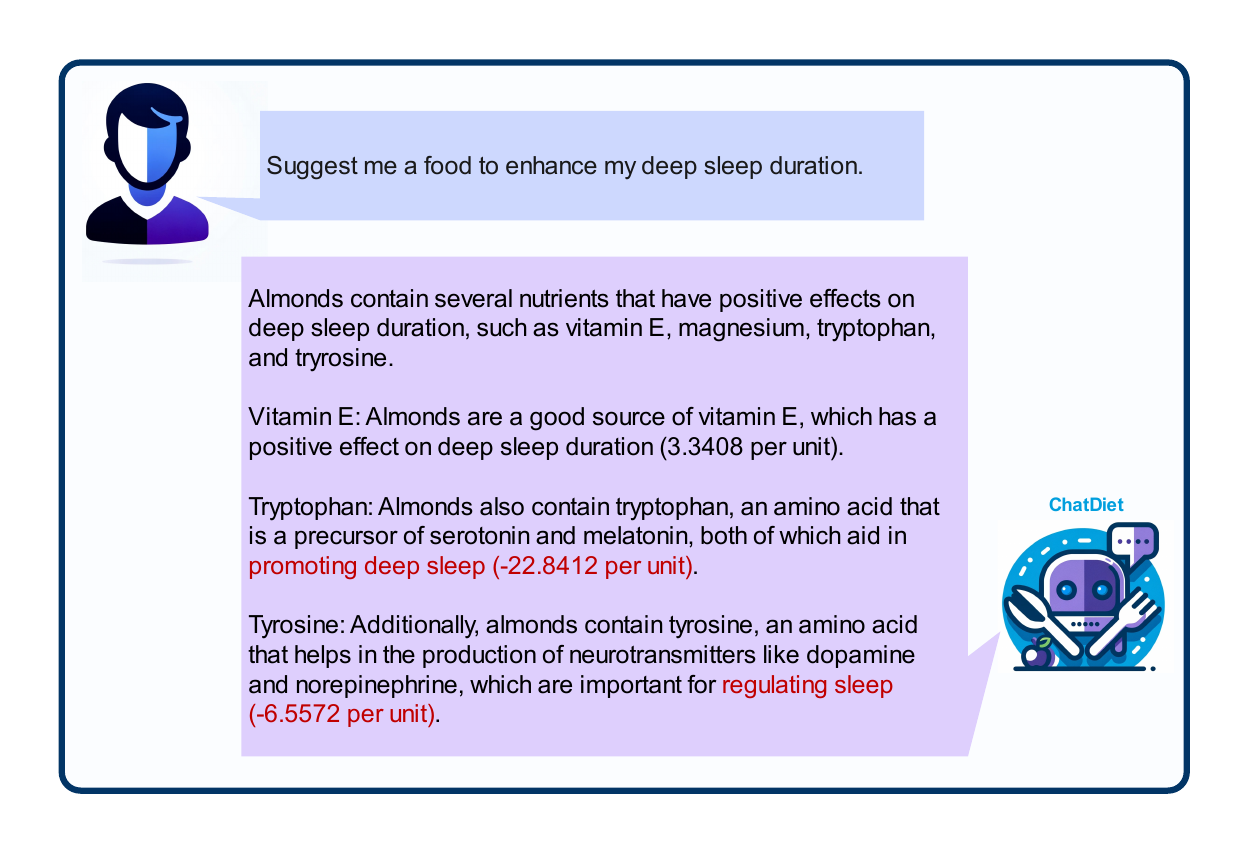}
    \caption{Hallucination: The nutrition effect statements are opposite to the shown causal effects.}
    \label{fig:selfcontra}
\end{figure}

One critical challenge we faced was hallucination in recommendations. In some cases, ChatDiet's outputs contradicted the provided personal nutrition effects, even when explicitly displayed. Figure~\ref{fig:selfcontra} exemplifies this issue. The language model acknowledges the personal impact by showing the calculated nutrition values. However, it then contradicts this by mentioning Tryptophan's supposed link to improved sleep, leading to an inaccurate recommendation for almonds. This self-contradiction undermines the credibility and reliability of ChatDiet's suggestions. This highlights the crucial need to address such inconsistencies and implement robust mechanisms to ensure the accuracy and coherence of ChatDiet's recommendations.

\section{Conclusion}

In this paper, we introduced ChatDiet, an LLM-powered framework for nutrition-oriented food recommendation. By combining Personal and Population Models with a dynamic Orchestrator, ChatDiet offered personalized and explainable food recommendations. Additionally, we implemented a chatbot based on ChatDiet using longitudinal data collected from an individual's dietary habits, health metrics, and other relevant information over three years. We further demonstrated the chatbot's practicality, achieving a 92\% effectiveness rate in food recommendations. 
Through dialogue examples, we showcased ChatDiet's strengths in explainability, personalization, and interactivity. ChatDiet represents a significant step in leveraging technology to improve dietary choices and overall well-being.

\section{Acknowledgment}
We extend our sincere gratitude to Tom Andriola, Vice Chancellor for Information Technology and Data and Chief Digital Officer at UC Irvine, for his invaluable contribution to this research through the provision of the data.
\bibliographystyle{unsrt}  
\bibliography{references,nitish_references}

\begin{thebibliography}{10}

\bibitem{mathers2019paving}
John~C Mathers.
\newblock Paving the way to better population health through personalised
  nutrition.
\newblock {\em EFSA Journal}, 17:e170713, 2019.

\bibitem{hu2002dietary}
Frank~B Hu.
\newblock Dietary pattern analysis: a new direction in nutritional
  epidemiology.
\newblock {\em Current opinion in lipidology}, 13(1):3--9, 2002.

\bibitem{raut2018personalized}
Madhu Raut, Keyur Prabhu, Rachita Fatehpuria, Shubham Bangar, and Sunita Sahu.
\newblock A personalized diet recommendation system using fuzzy ontology.
\newblock {\em Int. J. Eng. Sci. Invention}, 7(3):51--55, 2018.

\bibitem{Singh2017_gutMicrobiomeHumanHealth}
Rasnik~K. Singh, Hsin~Wen Chang, Di~Yan, Kristina~M. Lee, Derya Ucmak, Kirsten
  Wong, Michael Abrouk, Benjamin Farahnik, Mio Nakamura, Tian~Hao Zhu, Tina
  Bhutani, and Wilson Liao.
\newblock {Influence of diet on the gut microbiome and implications for human
  health}.
\newblock {\em Journal of Translational Medicine}, 15(1):1--17, 2017.

\bibitem{Valdes2018_GutMicrobiotaNutritionHealth}
Ana~M. Valdes, Jens Walter, Eran Segal, and Tim~D. Spector.
\newblock {Role of the gut microbiota in nutrition and health}.
\newblock {\em BMJ (Online)}, 361:36--44, 2018.

\bibitem{min2019food}
Weiqing Min, Shuqiang Jiang, and Ramesh Jain.
\newblock Food recommendation: Framework, existing solutions, and challenges.
\newblock {\em IEEE Transactions on Multimedia}, 22(10):2659--2671, 2019.

\bibitem{micha2017association}
Renata Micha, Jose~L Pe{\~n}alvo, Frederick Cudhea, Fumiaki Imamura, Colin~D
  Rehm, and Dariush Mozaffarian.
\newblock Association between dietary factors and mortality from heart disease,
  stroke, and type 2 diabetes in the united states.
\newblock {\em Jama}, 317(9):912--924, 2017.

\bibitem{story2008creating}
Mary Story, Karen~M Kaphingst, Ramona Robinson-O'Brien, and Karen Glanz.
\newblock Creating healthy food and eating environments: policy and
  environmental approaches.
\newblock {\em Annu. Rev. Public Health}, 29:253--272, 2008.

\bibitem{agapito2018dietos}
Giuseppe Agapito, Mariadelina Simeoni, Barbara Calabrese, Ilaria Car{\'e},
  Theodora Lamprinoudi, Pietro~H Guzzi, Arturo Pujia, Giorgio Fuiano, and Mario
  Cannataro.
\newblock Dietos: A dietary recommender system for chronic diseases monitoring
  and management.
\newblock {\em Computer methods and programs in biomedicine}, 153:93--104,
  2018.

\bibitem{iwendi2020realizing}
Celestine Iwendi, Suleman Khan, Joseph~Henry Anajemba, Ali~Kashif Bashir, and
  Fazal Noor.
\newblock Realizing an efficient iomt-assisted patient diet recommendation
  system through machine learning model.
\newblock {\em IEEE access}, 8:28462--28474, 2020.

\bibitem{jeon2023large}
Jaeho Jeon and Seongyong Lee.
\newblock Large language models in education: A focus on the complementary
  relationship between human teachers and chatgpt.
\newblock {\em Education and Information Technologies}, pages 1--20, 2023.

\bibitem{birkun2023large}
Alexei~A Birkun and Adhish Gautam.
\newblock Large language model-based chatbot as a source of advice on first aid
  in heart attack.
\newblock {\em Current Problems in Cardiology}, page 102048, 2023.

\bibitem{sapri2019diet}
Nurul Syafiqa~Mohd Sapri, Syariza Abdul-Rahman, Aida~Mauziah Benjamin, et~al.
\newblock A diet recommendation for diabetic patients using integer
  programming.
\newblock In {\em AIP Conference Proceedings}, volume 2138. AIP Publishing,
  2019.

\bibitem{zadeh2019personalized}
Maryam Sadat Amiri~Tehrani Zadeh, Juan Li, and Shadi Alian.
\newblock Personalized meal planning for diabetic patients using a
  multi-criteria decision-making approach.
\newblock In {\em 2019 IEEE International Conference on E-health Networking,
  Application \& Services (HealthCom)}, pages 1--6. IEEE, 2019.

\bibitem{toledo2019food}
Raciel~Yera Toledo, Ahmad~A Alzahrani, and Luis Martinez.
\newblock A food recommender system considering nutritional information and
  user preferences.
\newblock {\em IEEE Access}, 7:96695--96711, 2019.

\bibitem{elsweiler2015towards}
David Elsweiler and Morgan Harvey.
\newblock Towards automatic meal plan recommendations for balanced nutrition.
\newblock In {\em Proceedings of the 9th ACM Conference on Recommender
  Systems}, pages 313--316, 2015.

\bibitem{ng2017personalized}
Yiu-Kai Ng and Meilan Jin.
\newblock Personalized recipe recommendations for toddlers based on nutrient
  intake and food preferences.
\newblock In {\em Proceedings of the 9th international conference on management
  of digital ecosystems}, pages 243--250, 2017.

\bibitem{cioara2018expert}
Tudor Cioara, Ionut Anghel, Ioan Salomie, Lina Barakat, Simon Miles, Dianne
  Reidlinger, Adel Taweel, Ciprian Dobre, and Florin Pop.
\newblock Expert system for nutrition care process of older adults.
\newblock {\em Future Generation Computer Systems}, 80:368--383, 2018.

\bibitem{taweel2016service}
Adel Taweel, Lina Barakat, Simon Miles, Tudor Cioara, Ionut Anghel,
  Abdel-Rahman~H Tawil, and Ioan Salomie.
\newblock A service-based system for malnutrition prevention and
  self-management.
\newblock {\em Computer Standards \& Interfaces}, 48:225--233, 2016.

\bibitem{ge2015health}
Mouzhi Ge, Francesco Ricci, and David Massimo.
\newblock Health-aware food recommender system.
\newblock In {\em Proceedings of the 9th ACM Conference on Recommender
  Systems}, pages 333--334, 2015.

\bibitem{Min2021_foodKGReview}
Weiqing Min, Chunlin Liu, Leyi Xu, and Shuqiang Jiang.
\newblock {The Development and Applications of Food Knowledge Graphs in the
  Food Science and Industry}.
\newblock pages 1--45, 2021.

\bibitem{Gharibi2020_foodKG_ML}
Mohamed Gharibi, Arun Zachariah, and Praveen Rao.
\newblock {FoodKG: A Tool to Enrich Knowledge Graphs Using Machine Learning
  Techniques}.
\newblock {\em Frontiers in Big Data}, 3(April):1--12, 2020.

\bibitem{Sikka2019LearningIFood2019Challenge}
Karan Sikka.
\newblock {Learning User Preferences from Social Multimedia Analysis and
  Overview of the IFood2019Challenge}.
\newblock In {\em Proceedings of the 5th International Workshop on Multimedia
  Assisted Dietary Management}, MADiMa '19, page~18, New York, NY, USA, 2019.
  Association for Computing Machinery.

\bibitem{cui2022m6rec}
Zeyu Cui, Jianxin Ma, Chang Zhou, Jingren Zhou, and Hongxia Yang.
\newblock M6-rec: Generative pretrained language models are open-ended
  recommender systems, 2022.

\bibitem{huang2023chatgpt}
Hanyao Huang, Ou~Zheng, Dongdong Wang, Jiayi Yin, Zijin Wang, Shengxuan Ding,
  Heng Yin, Chuan Xu, Renjie Yang, Qian Zheng, et~al.
\newblock Chatgpt for shaping the future of dentistry: the potential of
  multi-modal large language model.
\newblock {\em International Journal of Oral Science}, 15(1):29, 2023.

\bibitem{liu2023pre}
Pengfei Liu, Weizhe Yuan, Jinlan Fu, Zhengbao Jiang, Hiroaki Hayashi, and
  Graham Neubig.
\newblock Pre-train, prompt, and predict: A systematic survey of prompting
  methods in natural language processing.
\newblock {\em ACM Computing Surveys}, 55(9):1--35, 2023.

\bibitem{zhang2021language}
Yuhui Zhang, Hao Ding, Zeren Shui, Yifei Ma, James Zou, Anoop Deoras, and Hao
  Wang.
\newblock Language models as recommender systems: Evaluations and limitations.
\newblock 2021.

\bibitem{liu2023chatgpt}
Junling Liu, Chao Liu, Renjie Lv, Kang Zhou, and Yan Zhang.
\newblock Is chatgpt a good recommender? a preliminary study.
\newblock {\em arXiv preprint arXiv:2304.10149}, 2023.

\bibitem{wang2023zero}
Lei Wang and Ee-Peng Lim.
\newblock Zero-shot next-item recommendation using large pretrained language
  models.
\newblock {\em arXiv preprint arXiv:2304.03153}, 2023.

\bibitem{geng2022recommendation}
Shijie Geng, Shuchang Liu, Zuohui Fu, Yingqiang Ge, and Yongfeng Zhang.
\newblock Recommendation as language processing (rlp): A unified pretrain,
  personalized prompt \& predict paradigm (p5).
\newblock In {\em Proceedings of the 16th ACM Conference on Recommender
  Systems}, pages 299--315, 2022.

\bibitem{dai2023uncovering}
Sunhao Dai, Ninglu Shao, Haiyuan Zhao, Weijie Yu, Zihua Si, Chen Xu, Zhongxiang
  Sun, Xiao Zhang, and Jun Xu.
\newblock Uncovering chatgpt's capabilities in recommender systems.
\newblock {\em arXiv preprint arXiv:2305.02182}, 2023.

\bibitem{lyu2023llm}
Hanjia Lyu, Song Jiang, Hanqing Zeng, Yinglong Xia, and Jiebo Luo.
\newblock Llm-rec: Personalized recommendation via prompting large language
  models.
\newblock {\em arXiv preprint arXiv:2307.15780}, 2023.

\bibitem{bao2023tallrec}
Keqin Bao, Jizhi Zhang, Yang Zhang, Wenjie Wang, Fuli Feng, and Xiangnan He.
\newblock Tallrec: An effective and efficient tuning framework to align large
  language model with recommendation.
\newblock {\em arXiv preprint arXiv:2305.00447}, 2023.

\bibitem{hou2023large}
Yupeng Hou, Junjie Zhang, Zihan Lin, Hongyu Lu, Ruobing Xie, Julian McAuley,
  and Wayne~Xin Zhao.
\newblock Large language models are zero-shot rankers for recommender systems.
\newblock {\em arXiv preprint arXiv:2305.08845}, 2023.

\bibitem{zhang2023recommendation}
Junjie Zhang, Ruobing Xie, Yupeng Hou, Wayne~Xin Zhao, Leyu Lin, and Ji-Rong
  Wen.
\newblock Recommendation as instruction following: A large language model
  empowered recommendation approach.
\newblock {\em arXiv preprint arXiv:2305.07001}, 2023.

\bibitem{li2023prompt}
Pan Li, Yuyan Wang, Ed~H Chi, and Minmin Chen.
\newblock Prompt tuning large language models on personalized aspect extraction
  for recommendations.
\newblock {\em arXiv preprint arXiv:2306.01475}, 2023.

\bibitem{lewis2021retrievalaugmented}
Patrick Lewis, Ethan Perez, Aleksandra Piktus, Fabio Petroni, Vladimir
  Karpukhin, Naman Goyal, Heinrich Küttler, Mike Lewis, Wen tau Yih, Tim
  Rocktäschel, Sebastian Riedel, and Douwe Kiela.
\newblock Retrieval-augmented generation for knowledge-intensive nlp tasks,
  2021.

\bibitem{oura}
Oura ring. smart ring for fitness, stress, sleep and health., Accessed March
  2024.
\newblock \url{https://ouraring.com/}.

\bibitem{mehrabadi2020sleep}
Milad~Asgari Mehrabadi, Iman Azimi, Fatemeh Sarhaddi, Anna Axelin, Hannakaisa
  Niela-Vil{\'e}n, Saana Myllyntausta, Sari Stenholm, Nikil Dutt, Pasi
  Liljeberg, Amir~M Rahmani, et~al.
\newblock Sleep tracking of a commercially available smart ring and smartwatch
  against medical-grade actigraphy in everyday settings: instrument validation
  study.
\newblock {\em JMIR mHealth and uHealth}, 8(11):e20465, 2020.

\bibitem{niela2022comparison}
Hannakaisa Niela-Vilen, Iman Azimi, Kristin Suorsa, Fatemeh Sarhaddi, Sari
  Stenholm, Pasi Liljeberg, Amir~M Rahmani, and Anna Axelin.
\newblock Comparison of oura smart ring against actigraph accelerometer for
  measurement of physical activity and sedentary time in a free-living context.
\newblock {\em CIN: Computers, Informatics, Nursing}, 2022.

\bibitem{cao2022accuracy}
Rui Cao, Iman Azimi, Fatemeh Sarhaddi, Hannakaisa Niela-Vilen, Anna Axelin,
  Pasi Liljeberg, and Amir~M Rahmani.
\newblock Accuracy assessment of oura ring nocturnal heart rate and heart rate
  variability in comparison with electrocardiography in time and frequency
  domains: comprehensive analysis.
\newblock {\em Journal of Medical Internet Research}, 24(1):e27487, 2022.

\bibitem{arboMadhu2018Personalized}
Arboleaf is to promote a healthier lifestyle in today's smart age, Accessed
  March 2024.
\newblock \url{https://www.arboleaf.com/}.

\bibitem{cronom}
Cronometer: Eat smarter. live better, Accessed March 2024.
\newblock \url{https://cronometer.com}.

\bibitem{applhealth}
Healthkit, Accessed March 2024.
\newblock \url{ https://developer.apple.com/health-fitness/}.

\bibitem{abbasian2023conversational}
Mahyar Abbasian, Iman Azimi, Amir~M Rahmani, and Ramesh Jain.
\newblock Conversational health agents: A personalized llm-powered agent
  framework.
\newblock {\em arXiv preprint arXiv:2310.02374}, 2023.

\bibitem{robertson2009probabilistic}
Stephen Robertson, Hugo Zaragoza, et~al.
\newblock The probabilistic relevance framework: Bm25 and beyond.
\newblock {\em Foundations and Trends{\textregistered} in Information
  Retrieval}, 3(4):333--389, 2009.

\bibitem{wei2022chain}
Jason Wei, Xuezhi Wang, Dale Schuurmans, Maarten Bosma, Fei Xia, Ed~Chi, Quoc~V
  Le, Denny Zhou, et~al.
\newblock Chain-of-thought prompting elicits reasoning in large language
  models.
\newblock {\em Advances in Neural Information Processing Systems},
  35:24824--24837, 2022.

\bibitem{rostami2020personal}
Ali Rostami, Vaibhav Pandey, Nitish Nag, Vesper Wang, and Ramesh Jain.
\newblock Personal food model.
\newblock In {\em Proceedings of the 28th ACM International Conference on
  Multimedia}, pages 4416--4424, 2020.

\bibitem{feng2023causal}
Qi~Feng.
\newblock Causal inference in diet, nutrition and health outcomes.
\newblock {\em Frontiers in Nutrition}, 10:1204695, 2023.

\bibitem{kalainathan2022structural}
Diviyan Kalainathan, Olivier Goudet, Isabelle Guyon, David Lopez-Paz, and
  Mich{\`e}le Sebag.
\newblock Structural agnostic modeling: Adversarial learning of causal graphs.
\newblock {\em The Journal of Machine Learning Research}, 23(1):9831--9892,
  2022.

\bibitem{shalit2017estimating}
Uri Shalit, Fredrik~D Johansson, and David Sontag.
\newblock Estimating individual treatment effect: generalization bounds and
  algorithms.
\newblock In {\em International conference on machine learning}, pages
  3076--3085. PMLR, 2017.

\bibitem{yao2021survey}
Liuyi Yao, Zhixuan Chu, Sheng Li, Yaliang Li, Jing Gao, and Aidong Zhang.
\newblock A survey on causal inference.
\newblock {\em ACM Transactions on Knowledge Discovery from Data (TKDD)},
  15(5):1--46, 2021.

\bibitem{nagesh2023towards}
Nitish Nagesh, Iman Azimi, Tom Andriola, Amir~M Rahmani, and Ramesh Jain.
\newblock Towards deep personal lifestyle models using multimodal n-of-1 data.
\newblock In {\em International Conference on Multimedia Modeling}, pages
  589--600. Springer, 2023.

\bibitem{pearl2014interpretation}
Judea Pearl.
\newblock Interpretation and identification of causal mediation.
\newblock {\em Psychological methods}, 19(4):459, 2014.

\bibitem{pearl2012mediation}
Judea Pearl.
\newblock The mediation formula: A guide to the assessment of causal pathways
  in nonlinear models.
\newblock {\em Causality: Statistical perspectives and applications}, pages
  151--179, 2012.

\bibitem{kalainathan2019causal}
Diviyan Kalainathan and Olivier Goudet.
\newblock Causal discovery toolbox: Uncover causal relationships in python,
  2019.

\bibitem{sharma2020dowhy}
Amit Sharma and Emre Kiciman.
\newblock Dowhy: An end-to-end library for causal inference, 2020.

\bibitem{haussmann2019foodkg}
Steven Haussmann, Oshani Seneviratne, Yu~Chen, Yarden Ne’eman, James Codella,
  Ching-Hua Chen, Deborah~L McGuinness, and Mohammed~J Zaki.
\newblock Foodkg: a semantics-driven knowledge graph for food recommendation.
\newblock In {\em The Semantic Web--ISWC 2019: 18th International Semantic Web
  Conference, Auckland, New Zealand, October 26--30, 2019, Proceedings, Part II
  18}, pages 146--162. Springer, 2019.

\bibitem{zeraatkar2019evidence}
Dena Zeraatkar, Bradley~C Johnston, and Gordon Guyatt.
\newblock Evidence collection and evaluation for the development of dietary
  guidelines and public policy on nutrition.
\newblock {\em Annual Review of Nutrition}, 39:227--247, 2019.

\bibitem{rossi2023alignment}
Laura Rossi, Marika Ferrari, and Andrea Ghiselli.
\newblock The alignment of recommendations of dietary guidelines with
  sustainability aspects: Lessons learned from italy’s example and proposals
  for future development.
\newblock {\em Nutrients}, 15(3):542, 2023.

\bibitem{cowan2023narrative}
Alexandra~E Cowan, Shinyoung Jun, Janet~A Tooze, Kevin~W Dodd, Jaime~J Gahche,
  Heather~A Eicher-Miller, Patricia~M Guenther, Johanna~T Dwyer, Nancy
  Potischman, Anindya Bhadra, et~al.
\newblock A narrative review of nutrient based indexes to assess diet quality
  and the proposed total nutrient index that reflects total dietary exposures.
\newblock {\em Critical Reviews in Food Science and Nutrition},
  63(12):1722--1732, 2023.

\bibitem{nutritionx}
Nutritionix - largest verified nutrition database, Accessed March 2024.
\newblock \url{https://www.nutritionix.com/}.

\bibitem{brown2020language}
Tom Brown, Benjamin Mann, Nick Ryder, Melanie Subbiah, Jared~D Kaplan, Prafulla
  Dhariwal, Arvind Neelakantan, Pranav Shyam, Girish Sastry, Amanda Askell,
  et~al.
\newblock Language models are few-shot learners.
\newblock {\em Advances in neural information processing systems},
  33:1877--1901, 2020.

\bibitem{ChatDietDemoVideo}
Chatdiet demo video, Accessed March 2024.
\newblock \url{https://www.youtube.com/watch?v=c-7IEBaRSyQ}.

\end{thebibliography}

\section{Appendix}
\begin{figure*}[ht]
    \centering
    \includegraphics[width=0.99\textwidth]{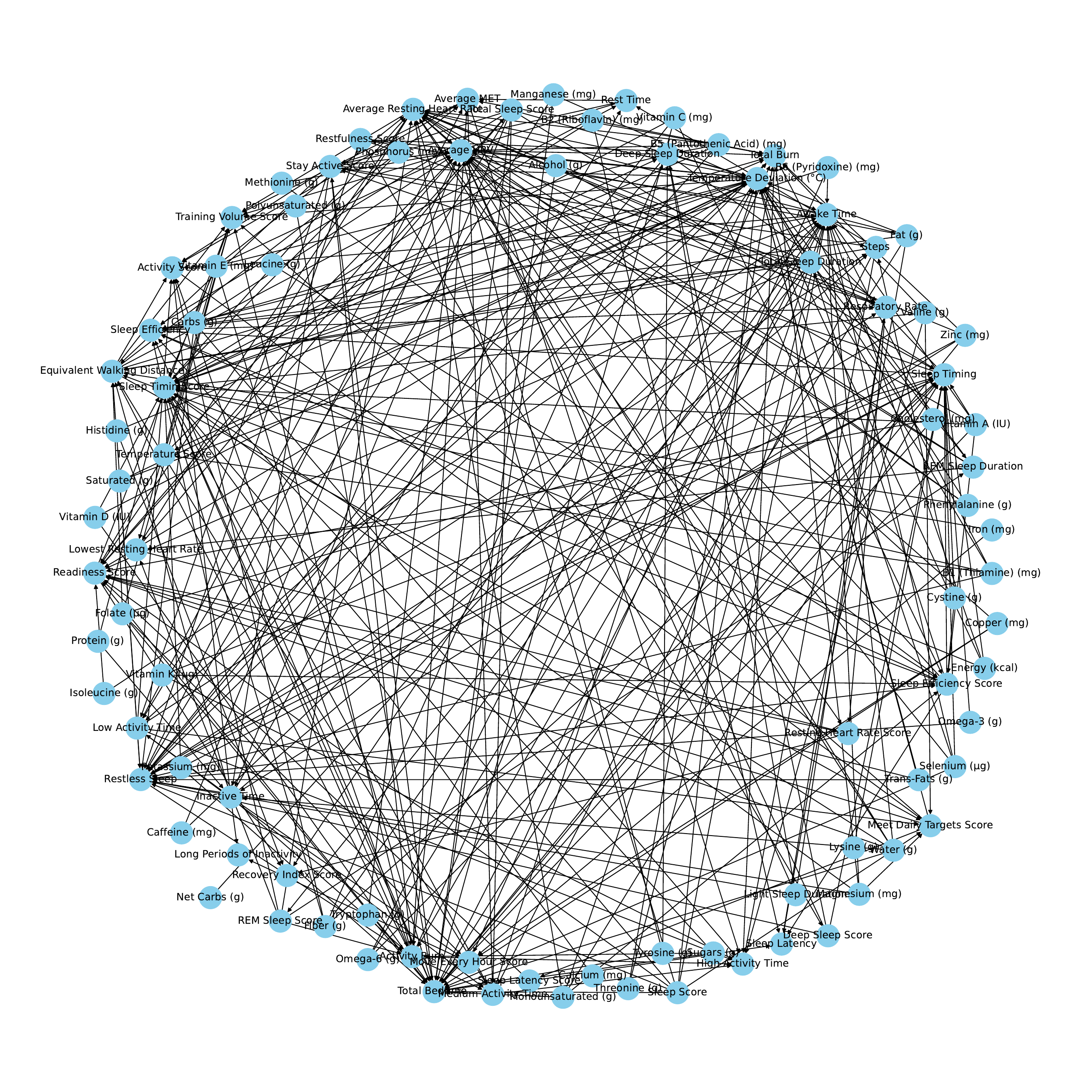}
    \caption{The Personal Nutrition Effect Causal Graph from the Personal Model}
        \centering
    \label{fig:causalgraph}
\end{figure*}

Figure~\ref{fig:causalgraph} illustrates the estimated causal graph between nutrition and health outcomes derived from the N-of-1 data.

\end{document}